\documentclass[twocolumn]{emulateapj}
\submitted{Science with Parkes @ 50 Years Young, 31 Oct. -- 4 Nov., 2011}

\usepackage{graphicx}

\shorttitle{The Parkes Pulsar Timing Array}
\shortauthors{Hobbs, G.}

\begin{document}


\title{The Parkes Pulsar Timing Array: \\
    What we've done and what we're doing}


\author{G. Hobbs}
\affil{CSIRO Astronomy and Space Science, Australia Telescope National Facility, PO~Box~76, Epping NSW~1710, Australia }



\begin{abstract}

First observations for the Parkes Pulsar Timing Array project were carried out in February 2004.  The project is ongoing and we currently observe approximately every three weeks. The data have led to numerous scientific results on topics as diverse as the solar wind, gravitational waves, measuring the masses of planetary systems in our solar system, atomic time scales, the interstellar medium and the pulsar emission mechanism.  In this paper we provide an historical overview of the project and highlight the major discoveries.

\end{abstract}





\section{Introduction}

For the past two decades the Parkes telescope has been  monitoring millisecond pulsars.  These observations have been used for numerous purposes including testing the general theory of relativity (e.g., van Straten et al. 2001) and for studying the pulse emission mechanism (e.g., Ord et al. 2004). Even though multiple pulsars were observed, these previous projects treated each pulsar individually.   

The first suggestions for a pulsar timing array (PTA) came from the North American community (Romani 1989; Foster \& Backer 1990).  Their work showed that the effects of some physical phenomena could be unambiguously identified by looking for correlated signals between different pulsar data sets.  They claimed that any identical signal in two or more pulsar data sets must correspond to a terrestrial phenomenon such as timing errors in the observatory instrumentation or irregularities in atomic time standards.  Partially correlated signals occur if the solar system planetary masses are not accurately known (Champion et al. 2010) or if gravitational waves (GWs) pass the Earth.  Detweiler (1979) had previously shown that pulsar observations were sensitive to GWs with nano-Hertz frequencies. Such GWs could be from individual supermassive binary black holes (Lommen \& Backer 2001), burst events (e.g., Finn \& Lommen 2010) or from a GW background formed from the incoherent addition of numerous GW sources (Hellings \& Downs 1983).  

In 2004, R. N. Manchester initiated the Parkes Pulsar Timing Array (PPTA) project (Manchester 2008).  The main goals were (and still are) to detect the GW background, improve our knowledge of solar system dynamics, produce a pulsar-based timescale and implement methods for radio frequency interference rejection.  In this paper we describe the history of the project and its current status. 

\section{The past}

The first request for allocated observing time was submitted in late 2003.  This proposal was written by four researchers at CSIRO and three from Swinburne University of Technology.  The first observation took place on the 6th of February 2004.   At that time, the basic parameters for a viable PTA were understood;  Foster \& Backer (1990) had indicated that at least five pulsars needed to be observed.  However, the necessary timing precision and the total data span required to produce scientifically-valuable data sets was unknown.  The only publically available observations of millisecond pulsars were from Kaspi et al. (1994). They had provided Arecibo Observatory observations of PSRs~J1857$+$0943 and J1939$+$2134 over a period of about 8\,yr.  PSR~J1857$+$0943 had rms timing residuals $\sim 1 \mu$s and the timing of J1939$+$2134 was dominated by an unexplained red-noise process. Upper bounds on the existence of a GW background with characteristics expected from an inflationary model of the early universe were known, and generally presented as a limit on the energy density, per logarithmic frequency interval, in a cosmic background of GWs, but predictions from Jaffe \& Backer (2003), Wyithe \& Loeb (2003) and Caldwell \& Allen (1992) were showing that a GW background formed by coalescing supermassive binary black holes at the cores of merging galaxies or a background from cosmic strings was more likely.  

Jenet et al. (2004) provided the first major astrophysical result using both pulsars and GWs when they constrained the properties of a possible supermassive binary black hole system within the 3C66B radio galaxy.  However, this result only made use of one pulsar and so did not relate directly to the PTA concept. A few papers had been written on developing a pulsar-based timescale (e.g., Guinot \& Petit 1991; Petit \& Tavella 1996), but no extensive attempt had been made to do this.  Very little literature was available describing how pulsar observations could improve our knowledge of the solar system. 

\begin{figure*}
\begin{center}
\includegraphics[width=15.5cm]{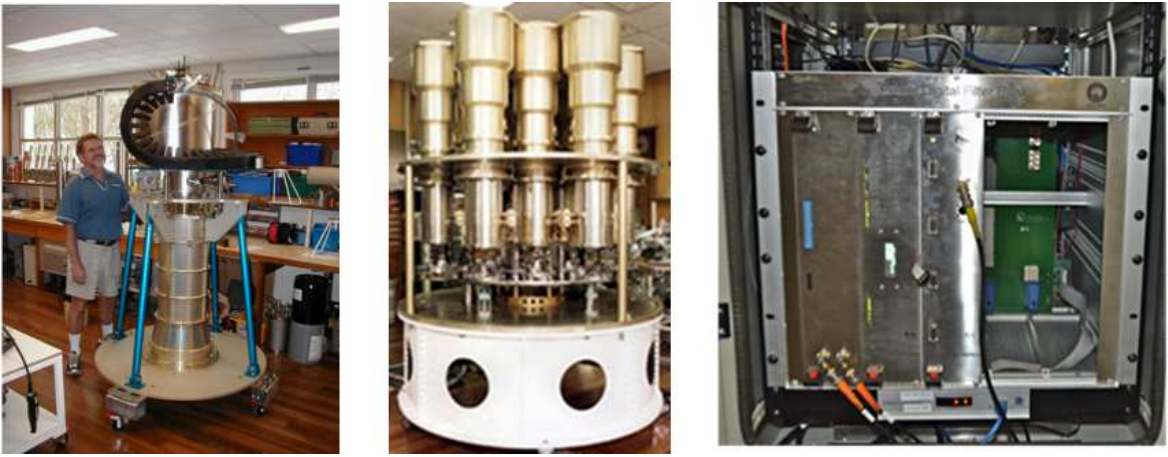}
\end{center}
\caption{The basic observing system at the Parkes telescope.  The left figure shows the 10-50\,cm dual band receiver.  The central panel contains the 20\,cm multibeam system and the right panel shows a digital filterbank backend system.}\label{fg:pks}
\end{figure*}

At the Parkes Observatory, various front-end systems and backend instruments were available.  The main front-end receiver systems for pulsar timing observations included the 20\,cm multibeam receiver (Stavely-Smith et al. 1996), the H-OH receiver and a dual-band 10/50\,cm receiver (Granet et al. 2005). Backend instruments included an analogue filterbank system, the ``wide-band-correlator" (WBC) and the Caltech-Swinburne-Parkes-Recorder 2 (CPSR2).  CPSR2 provided two 64\,MHz bands of coherently dedispersed data whereas the WBC provided a wider bandwidth (up to 512\,MHz), but the data were not coherently dedispersed.  The standard observing receivers (the 20\,cm multibeam and the dual-band 10/50\,cm receivers) are shown in Figure~\ref{fg:pks} along with a recent backend instrument (the Parkes Digital Filterbank).

\subsection{2005 and 2006}

\begin{figure*}
\begin{center}
\includegraphics[angle=-90,width=15.5cm]{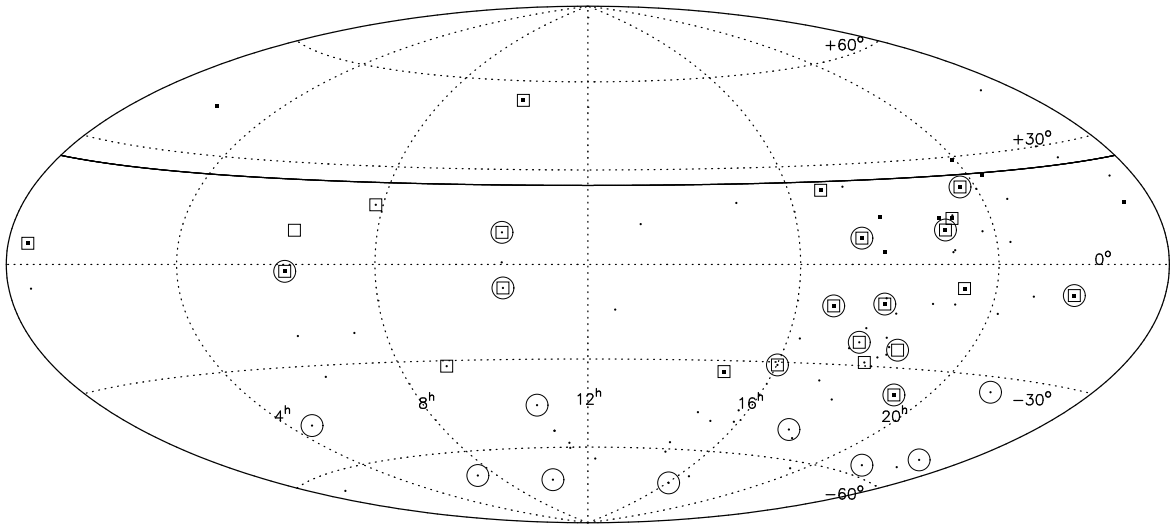}
\end{center}
\caption{Sky-map of the pulsars included in the Parkes Pulsar Timing Array sample (large open circles), the European Pulsar Timing Array (open boxes), NANOGrav (solid boxes) and all known pulsars with $P < 20$\,ms, $\dot{P} < 10^{-18}$ that do not reside in a globular cluster (small dots). The solid line at a declination of $+25^\circ$ gives the approximate declination limit of the Parkes Observatory.}\label{fg:skymap}
\end{figure*}

The PPTA project started to produce high-quality observations in March 2005.  We selected around 20 millisecond pulsars based on their pulse periods, known stability and flux density.  This number was based on the analysis of Jenet et al. (2005) who demonstrated that an isotropic, stochastic GW background with an expected amplitude could be detected with $>3\sigma$ confidence if $\sim$20 pulsars were timed weekly over a period of five years with an rms timing residual of 100\,ns.  A map of the pulsar positions on the sky is given in Figure~\ref{fg:skymap}.  The Figure includes 22 PPTA pulsars.  Originally, a slightly larger sample of pulsars were chosen, but PSRs~J0024$-$7204J, J1435$-$6100, J1623$-$2631, J1721$-$2457, J1757$-$5322, J2317$+$1439 did not provide adequate timing precision and were quickly dropped.  More recently PSR~J1732$-$5049 was also removed, but, as described below, two recently discovered pulsars have now been added into the sample.

Attempting to achieve the required pulse time of arrival (ToA) precision for such a large sample of pulsars was (and still is) very ambitious.  Van Straten (2001) had showed that rms timing residuals of $<$100\,ns were possible for a single pulsar, PSR~J0437$-$4715, but such timing precision had not been published for any other pulsar.  Obviously we would need new hardware to achieve the required level.  Planning therefore begun on designing a new set of instrumentation. During June 2005 we developed a test system, the first Parkes Digital Filterbank (PDFB1) that, along with CSPR2, provided the first major data set for the PPTA project.   Interference caused by digital television transmissions led to Kesteven et al. (2005) developing adaptive filter algorithms that could be used to mitigate such interference in our observations. 

The first major result from the PPTA project was published by Jenet et al. (2006).  This provided the most stringent upper bound on the existence of the GW background.  This work, and most subsequent PPTA papers, have been based on a new pulsar software package, \textsc{tempo2} (Hobbs, Edwards \& Manchester 2006; Edwards, Hobbs \& Manchester 2006).  Using this new software we can now 1) process multiple pulsars simultaneously, 2) simplify new algorithmic development and 3) implement all the relevant physical processes sufficiently accurately for our project goals.


\subsection{2007 and 2008}

The timing residuals induced by a GW background, irregularities in terrestrial time standards or errors in the solar system ephemeris are all expected to have a steep low-frequency (``red") spectrum.  The timing residuals induced by dispersion measure variations are also predicted to have a red noise spectrum that, for many pulsars, will be larger than the signals that we hope to detect.  You et al. (2007a,b) measured the size of the dispersion measure variations caused by the interstellar medium and the solar wind.  This work highlighted the necessity for repeated observations using the 10/50cm dual-band feed which requires two backend instruments for data capture and processing.  For the low frequency band it is necessary to  de-disperse the observations coherently.  Therefore CPSR2 was used for observations in the 50\,cm band whereas the PDFB1 was used at the higher frequency. PDFB1 could only process a bandwidth of 256\,MHz.  A second digital filterbank system (PDFB2), able  to process the entire available bandwidth of $\sim$1024\,MHz, was commissioned in March 2007.

\begin{figure*}
\begin{center}
\includegraphics[width=13cm]{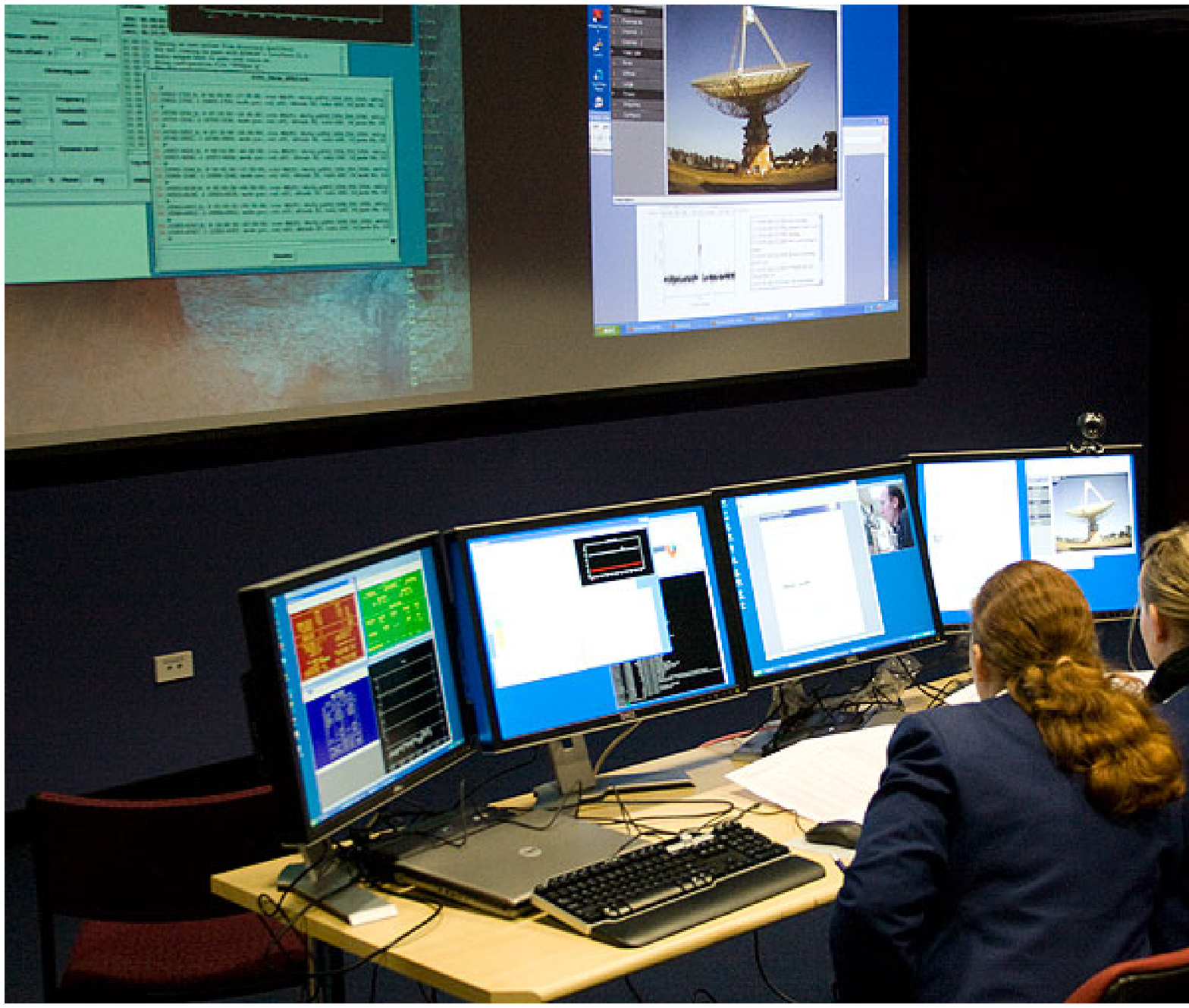}
\end{center}
\caption{High school students from Caroline Chisholm College observing pulsars as part of the PULSE@Parkes outreach project. Photo credit: CSIRO.}\label{fg:patp}
\end{figure*}

Any project related to pulsars, black holes, galaxies, clocks and planets has the potential to engage the public and school students.  In December 2004 students from Kingswood High School (in New South Wales,  Australia) were the first to take part in a PULSE@Parkes observing session (Hollow et al. 2008, Hobbs et al. 2009a).  This outreach project allows high school students to carry out observations of pulsars using the Parkes telescope remotely (Figure~\ref{fg:patp}).  Their sample of pulsars includes many of the PPTA pulsars providing new observations while also testing out the remote observing capabilities at the Observatory that will soon become mainstream.

\subsection{2009 and 2010}

In order to develop algorithms to detect GW signals or provide an upper bound on their amplitude it is necessary to predict the size of the induced timing residuals that will be present for a specific pulsar.  Sesana et al. (2008, 2009) described the expected signal strength for individual sources and refined the calculations for a  background induced by a large number of sources.  Determining the size of the induced timing residuals is not trivial as the pulsar timing process requires the fitting and subsequent removal of various analytic functional forms.  This can reduce the size of any GW signal in the timing residuals.   Hobbs et al. (2009b) developed a set of routines that allow the expected GW signal to be simulated within the \textsc{tempo2} software package.  This allowed the effects of fitting to be accounted for when predicting the GW signal.  The first analysis of the sensitivity of  an actual PTA to individual, continuous sources of GWs was presented by Yardley et al. (2010).  This led to a sky-averaged constraint on the merger rate of nearby ($z < 0.6$) black hole binaries with a chirp mass of $10^{10}$\,M$_\odot$ of less than one merger every seven years.

Champion et al. (2010) demonstrated progress towards another of the PPTA goals: looking for errors in the planetary ephemeris.  We enhanced the \textsc{tempo2} software to enable a signal to be fitted to multiple pulsars simultaneously.  This allowed errors in the masses of known solar system planets to be identified and led to the most precise published estimate for the mass of the Jovian system of $9.547921(2)\times 10^{-4}$\,M$_\odot$.

It was becoming clear that the timing residuals for many of the pulsars in the PPTA sample are dominated by an unknown red noise process that may be similar to the ``timing noise'' observed for the normal pulsars.  An analysis of 366 pulsars using data from the Jodrell Bank Observatory allowed Hobbs, Lyne \& Kramer (2010) to compare various mechanisms for this noise and to predict the expected amount of noise for millisecond pulsars.  This was further developed by Lyne et al. (2010) who identified a process that led to variations in pulse shape that correlated with the spin-down rate.  This opened up the possibility that the timing noise in millisecond pulsars could be measured and subsequently removed.  Much of the low-frequency noise in our long data sets had previously been absorbed into arbitrary time offsets that occurred whenever the observing systems changed.    We implemented a new procedure to measure the time delay between the topocentric reference point of the telescope and the time tagging on the output data for as many of our backend systems as possible.  This, and correlations of overlapping data sets, enabled the removal of all such arbitrary time offsets within the PPTA data set.

Verbiest et al. (2009) provided new analytic calculations that predicted the likely GW background detection significance for a given set of observations. Their results showed the necessity of observing even more than 20 pulsars in order to detect the GW background within five years.  In order to produce data sets with a sufficiently large number of pulsars, three PTA projects, the PPTA, the European Pulsar Timing Array (EPTA) and the North American Timing Array (NANOGrav) agreed to collaborate and share data sets.  The resulting International Pulsar Timing Array (IPTA) has been described by Hobbs et al. (2010).  

During this time Keith et al. (2010) presented the first results from the ``High Time Resolution Universe Pulsar Survey" being carried out at the Parkes Observatory.  This survey is discovering new millisecond pulsars some of which are now being included in the IPTA pulsar sample.  For instance, in September 2010 we added a newly discovered pulsar, PSR~J1017$-$7156, to the PPTA sample.  Searches for radio pulsars using gamma-ray source catalogues has also recently led to the discovery of a large number of millisecond pulsars.  Regular observations of a newly discovered pulsar  from this survey, PSR~J2241$-$5236 (Keith et al. 2011), began in February 2010.

In June 2010, the coherent de-dispersion system (CPSR2) was decommissioned and replaced by the ATNF-Parkes-Swinburne-Recorder (APSR) that enabled coherent dedispersion across a bandwidth of 1\,GHz. By October 2010 we were making use of a second coherent de-dispersion backend system, CASPSR,  in parallel with APSR.




\subsection{2011 and current research}

The PPTA upper bounds that had already been placed on GW signals (both individual sources and the GW background) were used by Wen et al. (2011) to constrain the coalescence rate of supermassive black-hole binaries, but still no detection of the GWs had been made.  In order to attempt such a detection, Yardley et al. (2011) developed new algorithms that were able to detect the GW background in simulated data.  This work showed that our observations were consistent with the hypothesis that no GW background is present with 76\% confidence. The Yardley et al. algorithm  is not effective in the presence of significant low-frequency noise. This led to a new research paper, published by Coles et al. (2011), describing how pulsar data should be analysed in the presence of red noise.   We can use these new methods for standard pulsar timing experiments as well as for PTA-related research.   

The first stage in any analysis of the PPTA observations relies on forming high-quality, calibrated pulse profiles.  Profiles from PDFB2 in the 20\,cm observing band were published by Yan et al. (2011a,b) who identified many new features.  Any time variability in the pulse profiles lead to an extra high frequency noise component. This was studied in detail by Oslowski et al. (2011) using 25\,h of observations of PSR~J0437$-$4715. They showed that the ToA precision achievable will ultimately be limited by the broad-band intensity modulation that is intrinsic to the pulsar's stochastic radio signal.

The raw observation data files produced for the PPTA project are large;  many terabytes of data now exist.  In order to provide public access to these data sets, we developed the Parkes Observatory Pulsar Data Archive (Hobbs et al. 2011).  This online archive includes as much recoverable data as possible from observations of pulsars at the Parkes Observatory and includes all the PPTA observations.  Raw data files are automatically converted to the PSRFITS format (Hotan et al. 2004) and are available for download 18 months after the data collection.   You et al. (2012) selected a small number of PPTA observations for PSR~J1022$+$1001 when the line of sight from the Earth to the pulsar went close to the Sun.  Using these observations they provided measurements of the solar wind electron density and its magnetic field strength.

The first major data release will soon be made and a full description of the PPTA project, its goals, the hardware used and its status will be reported in Manchester et al. (2012, in preparation).  The preparation of this data set has required the development of a new method for removing the effects of dispersion measure variations (Keith et al. 2012, in preparation).  The data have been used to form the first time scale based on the PPTA sample of pulsars (Hobbs et al. 2012, in preparation).   Furthermore we are developing new algorithms to limit the amplitude of the GW background and our expected new result (to be published by Shannon et al. 2012) will significantly improve upon the  limit (van Haasteren et al. 2011) obtained using EPTA data sets.  

\section{The future}

The development of a well-tested, optimal algorithm for detecting GW signals, a full understanding of all the phenomena that can affect our data sets and a prediction of the expected size of the GW signal are essential for the PPTA project.  Not only will these developments allow for the possibility of GW detection - the main, and most exciting goal of the PPTA project - but will also allow us to predict when this momentous detection is likely to occur.    It is unlikely that the sample of pulsars will significantly change in the next few years, but we shall doubtless include new discoveries in the PPTA sample.  It is also possible that new algorithms will be developed for the PPTA goals that make use of the orbital periodicity of binary pulsars instead of a pulsar's rotational periodicity (see e.g., Kopeikin, 1997; Ilyasov, Kopeikin \& Rodin 1998).  If this occurs we may need to include a selection of neutron-star--neutron-star binary systems in the PPTA sample.  

The PPTA data sets and algorithmic developments are being used in unexpected ways.  Recent theoretical work has indicated that PTA observations are sensitive to the gravitational wave memory effect (e.g., van Haasteren \& Levin 2010 and references therein); an effect unlikely to be detectable using other GW instruments.  The \textsc{tempo2} software we have developed has led to many unexpected research projects.  It is currently being used to determine how pulsars may help in the navigation of space probes in our solar system (e.g., Tian et al. 2012).  

It is likely that most future analyses will use IPTA observations that combine the PPTA, EPTA and NANOGrav data sets.   Within a decade it is expected that the Square Kilometre Array pathfinder telescopes, ASKAP (Johnston et al. 2007) and Meerkat will be producing data to be included in the IPTA.  The Chinese Five-hundred-metre Aperture Spherical Telescope (FAST) will provide a collecting area that should revolutionise pulsar astronomy.  Combined with a planned 110\,m diameter telescope at the Xinjiang Observatory, the Chinese community will have the data sets for a powerful new PTA.  One of only two key science projects for the Phase 1 Square Kilometre Array (SKA) is related to high precision pulsar timing and searching for GWs. It is therefore to be expected that GWs will have been detected before the commissioning of the full SKA telescope. The number of pulsars observable and the phenomenal sensitivity will allow the SKA to produce data sets that not only will be used to detect GWs, but will allow the GW properties to be studied in detail.

\section{Conclusion}

The first PPTA observing proposal, in 2003, contained seven Australian members.  The current team consists of 31 people across Australia, Germany, Poland, China and the U.S.A.  The viability of using pulsar observations to detect GWs has now been recognised by the wider GW community with the number of publications related to PTAs increasing dramatically over the previous five years.

Pulsar observations led to the first indirect detection of GWs using the PSR~B1913+16 pulsar system (Hulse \& Taylor 1975). With the significant ongoing world-wide effort, it is possible that PTAs will make the first direct detection of GWs.  Pulsar projects provide the only viable means to study ultra-low frequency GWs and so PTAs are complementary to ground-based GW detector systems such as LIGO and VIRGO.  GW detection is only one of the PPTA goals.  Using PPTA data, we have already developed the first major pulsar time scale as we continue to search for unknown objects  within our solar system.

\acknowledgments

The PPTA project was initiated with support from R. N. Manchester's Australian Research Council (ARC) Federation Fellowship (FF0348478).  We acknowledge contributions to the project Worldwide from numerous individuals and institutes.  GH acknowledges support from the Australian Research Council \#DP0878388.   The Parkes radio telescope is part of the Australia Telescope which is funded by the Commonwealth of Australia for operation as a National Facility managed by CSIRO.


\begin{references}

\reference{ca92}Caldwell, R. R., Allen, B., 1992, PhRvD, 45, 3447

\reference{cha10}Champion, D. J., et al., 2010, ApJ, 720, L201

\reference{col11}Coles, W., Hobbs, G., Champion, D. J., Manchester, R. N., Verbiest, J. P. W., 2011, MNRAS, 418, 561

\reference{det79}Detweiler, S., 1979, ApJ, 234, 1100

\reference{ehm06}Edwards, R., Hobbs, G., Manchester, R., 2006, MNRAS, 372, 1549

\reference{fl10}Finn, L. S. \& Lommen, A., 2010, ApJ, 718, 1400

\reference{fb90}Foster R. S., Backer, D. C., 1990, ApJ, 361, 300

\reference{gp1991}Guinot, B., Petit, G., 1991, A\&A, 248, 292

\reference{gra05}Granet et al., 2005, IEEE TAP, 47, 13

\reference{hd83}Hellings, R. W. \& Downs, G. S., 1983, ApJ, 265, L39

\reference{hem06}Hobbs, G., Edwards, R., Manchester, R., 2006, MNRAS, 369, 655

\reference{hob09}Hobbs, G., et al. 2009, PASA, 26, 468a

\reference{hob09}Hobbs, G., et al. 2009, MNRAS, 394, 1945b

\reference{hlk10}Hobbs, G., Lyne, A., Kramer, M., 2010, MNRAS, 402, 1027

\reference{hob10}Hobbs, G., et al., 2010, Classical and Quantum Gravity, 27, 4013

\reference{hob011}Hobbs, G., et al., 2011, PASA, 28, 202

\reference{hol08}Hollow, R. et al. 2008, ASPC, 400, 190

\reference{hob04}Hotan, A. W., van Straten, W., Manchester, R. N., 2004, 21, 302 

\reference{ht75}Hulse, R. \& Taylor, J., 1975, ApJ, 195, L51

\reference{ikr98}Ilyasov, Kopeikin, S., Rodin. A., 1998, AstL, 24, 228

\reference{jb03}Jaffe, A. H., Backer, D. C., 2003, ApJ, 583, 616

\reference{jllw04}Jenet, F., Lommen, A., Larson, S. L., Wen, L., ApJ, 2004, 606, 799

\reference{jen05}Jenet, F., et al. 2005, ApJ, 625, 123

\reference{jen06}Jenet, F., et al. 2006, ApJ, 653, 1571

\reference{joh07}Johnston, S., et al., 2007, PASA, 24, 174

\reference{ktr94}Kaspi, V., M., Taylor, J. H., Ryba, M., 1994, ApJ, 428, 713

\reference{kei10}Keith, M., et al., 2010, MNRAS, 409, 619

\reference{kei11}Keith, M., et al., 2011, MNRAS, 414, 1929

\reference{kes05}Kesteven, M., Hobbs, G., Clement, R., Dawson, B., Manchester, R., Uppal, T., 2005, RaSc, 40, 5S06

\reference{kop97}Kopeikin, S., 1997, PhRvD, 56, 4455

\reference{lb01}Lommen, A. \& Backer, D., 2001, ApJ, 562, 297

\reference{lyn10}Lyne, A., Hobbs, G., Kramer, M., Stairs, I., Stappers, B., 2010, Science, 329, 408

\reference{man08}Manchester, R. N., 2008, in {\it 40 Years of Pulsars: Millisecond Pulsars, Magnetars and More},  AIP Conf.Proc. 983, p.584

\reference{ord04}Ord, S. M., van Straten, W., Hotan, A. W., Bailes, M., 2004, \mnras, 353, 804

\reference{osl11}Oslowski, S., van Straten, W., Hobbs, G., Bailes, M., Demorest, P., 2011, MNRAS, 418, 1258

\reference{pt96}Petit, G., \& Tavella, P., 1996, A\&A, 308, 290

\reference{rom89}Romani, R. W., 1989, in {\it Timing Neutron Stars}, Eds: H. Ogelman and E.P.J. van den Heuvel, Kluwer Academic, p.113

\reference{ses08}Sesana, A., Vecchio, A., Colacino, C. N., 2008, MNRAS, 390, 192

\reference{ses09}Sesana, A., Vecchio, A., Volonteri, M., 2009, MNRAS, 394, 4, 2255

\reference{ss96}Staveley-Smith, L., et al., 1996, PASA, 13, 243

\reference{tian12}Tian, F., Tang, Z-H., Ya, Q-Z., Yu, Y., 2012, RAA, 12, 219

\reference{hl10}van Haasteren, R., \& Levin, Y., 2010, MNRAS 401, 2372

\reference{haa11}van Haasteren, R., et al., 2011, MNRAS, 414, 3117

\reference{vs01}van Straten, W. et al., 2001, Nature, 412, 158

\reference{ver09}Verbiest, J. P. et al., 2009, MNRAS, 400, 951

\reference{wen11}Wen, Z. L., Jenet, F. A., Yardley, D., Hobbs. G., Manchester, R. N., 2011, ApJ, 730, 29

\reference{wl03}Wyithe, S., Loeb, A., 2003, ApJ, 595, 614

\reference{yan11a}Yan, W. M., et al., 2011, MNRAS, 414, 2087a

\reference{yan11}Yan, W. M., et al., 2011, Ap\&SS, 335, 485b

\reference{yar10}Yardley, D., et al., 2010, MNRAS, 407, 669

\reference{yar11}Yardley, D., et al., 2011, MNRAS, 414, 1777

\reference{you07}You, X. P., et al., 2007, MNRAS, 378, 493a

\reference{you07b}You, X. P., et al., 2007, ApJ, 671, 907b

\reference{you12}You, X. P., et al., 2012, MNRAS, in press (arXiv, 1202, 2263)




\end{references}
\end{document}